\documentclass{article}
\usepackage{arxiv}
\usepackage{graphicx}
\usepackage{hyperref}
\usepackage{siunitx}
\usepackage{amssymb}
\usepackage{amsfonts}
\usepackage{threeparttable}
\usepackage{caption}
\usepackage{natbib}

\begin{document}

\title{Improved Protein-ligand Binding Affinity Prediction with Structure-Based Deep Fusion Inference}

\author{Derek Jones\,$^{\dagger}$\thanks{$\dagger$ denotes authors contributed equally} \\ Global Security Computing Applications Division \\ Lawrence Livermore National Laboratory\\Livermore, CA \And
Hyojin Kim\,$^{\dagger}$ \\ Center for Applied Scientific Computing\\Lawrence Livermore National Laboratory \\ Livermore, CA \And 
Xiaohua Zhang \\ Biosciences and Biotechnology Division \\ Lawrence Livermore National Laboratory \\ Livermore, CA \And 
Adam Zemla \\ Biosciences and Biotechnology Division \\ Lawrence Livermore National Laboratory \\ Livermore, CA \And
Garrett Stevenson \\ Computational Engineering Division \\ Lawrence Livermore National Laboratory \\ Livermore, CA \And
William D. Bennett \\ Biosciences and Biotechnology Division \\ Lawrence Livermore National Laboratory \\ Livermore, CA \And
Dan Kirshner \\ Biosciences and Biotechnology Division \\ Lawrence Livermore National Laboratory \\ Livermore, CA \And 
Sergio Wong \\ Biosciences and Biotechnology Division \\ Lawrence Livermore National Laboratory \\ Livermore, CA \And
Felice Lightstone \\ Biosciences and Biotechnology Division \\ Lawrence Livermore National Laboratory \\ Livermore, CA \And
Jonathan E. Allen \\ Global Security Computing Applications Division \\ Lawrence Livermore National Laboratory \\ Livermore, CA}

\maketitle

Predicting accurate protein-ligand binding affinity is important in drug discovery but remains a challenge even with computationally expensive biophysics-based energy scoring methods and state-of-the-art deep learning approaches. Despite the recent advances in the deep convolutional and graph neural network based approaches, the model performance depends on the input data representation and suffers from distinct limitations. It is natural to combine complementary features and their inference from the individual models for better predictions. We present fusion models to benefit from different feature representations of two neural network models to improve the binding affinity prediction.
We demonstrate effectiveness of the proposed approach by performing experiments with the PDBBind 2016 dataset and its docking pose complexes. The results show that the proposed approach improves the overall prediction compared to the individual neural network models with greater computational efficiency than related biophysics based energy scoring functions. We also discuss the benefit of the proposed fusion inference with several example complexes. 
The software is made available as open source at \url{https://github.com/llnl/fast}.

\section{Introduction}
Predicting accurate binding affinity between a small molecule and target protein is one of the fundamental challenges in drug development. Recently deep learning models have been proposed as an alternative to traditional physics-based free energy scoring functions. The benefit of the deep learning approach is in learning binding interaction rules directly from an atomic representation without relying on hand curated features that may not capture the mechanism of binding~(\citealt{ballester_machine_2010, ain_machine-learning_2015}). Two types of 3D structure based deep learning approaches are addressed in this literature, 3D convolutional neural networks (3D-CNN) and spatial graph convolutional neural network (SG-CNN) models~(\citealt{feinberg_potentialnet_2018, zhang_deepbindrg_2019}).  The 3D-CNN, similar to the one proposed in \citealt{ragoza_proteinligand_2017}, models the atoms with a 3D voxel representation.  The 3D-CNN representation implicitly models pairwise relationships between atoms through the relative positioning of atoms in a 3D voxel grid, but does not pre-determine, which atomic interactions to represent other than defining a minimum atomic resolution. This comes at the cost of having to learn a larger number of parameters to represent the grid.  In contrast, the SG-CNN uses an explicit distance threshold to determine, which pairs of atoms to consider in pairwise interactions, but with the potential benefit of using fewer parameters in the model. Both approaches show promise, with potential complementary strengths and weakness. However, both methods have yet to be compared directly with each other and relative to a traditional physics-based scoring function. Recently, there has been considerable success in employing fusion models for a variety of computer vision tasks such as video activity recognition and multi-modal image fusion. The benefit of the fusion models lies in combining different input modalities and learning their feature representations together.  Several fusion models have been proposed for the task of video activity recognition by bridging the dimension and feature difference among multiple input modalities (e.g., visual and temporal data) (\citealt{cheron_p-cnn_2015, huang_human_2019}). The work of \citealt{roitberg_analysis_2019} addressed several distinct strategies of fusion models and their performance differences for the task of gesture recognition in video. Similar ideas have also been applied to multi modal image fusion problems such as road detection from different sensors~(\citealt{yang_fusion_2018}). Inspired by the fusion models in computer vision applications, we introduce a fusion model designed to combine independently trained 3D-CNN and SG-CNN models, each of which is expected to capture different characteristics of atomic representations. To the best of our knowledge, the proposed fusion approach is the first attempt to combine heterogeneous model representations for the task of protein-ligand binding affinity prediction. We evaluate our approach on the crystal structure and docking pose complexes of the PDBBind 2016 dataset ~(\citealt{burley_protein_2019}). The results show that the fusion model can successfully combine complimentary predictions from the constituent models. Furthermore, this approach has the potential to be a more efficient alternative to more computationally expensive scoring functions while yielding improved prediction.

\begin{figure*}[ht]
  \centering
    \includegraphics[width=0.75\linewidth]{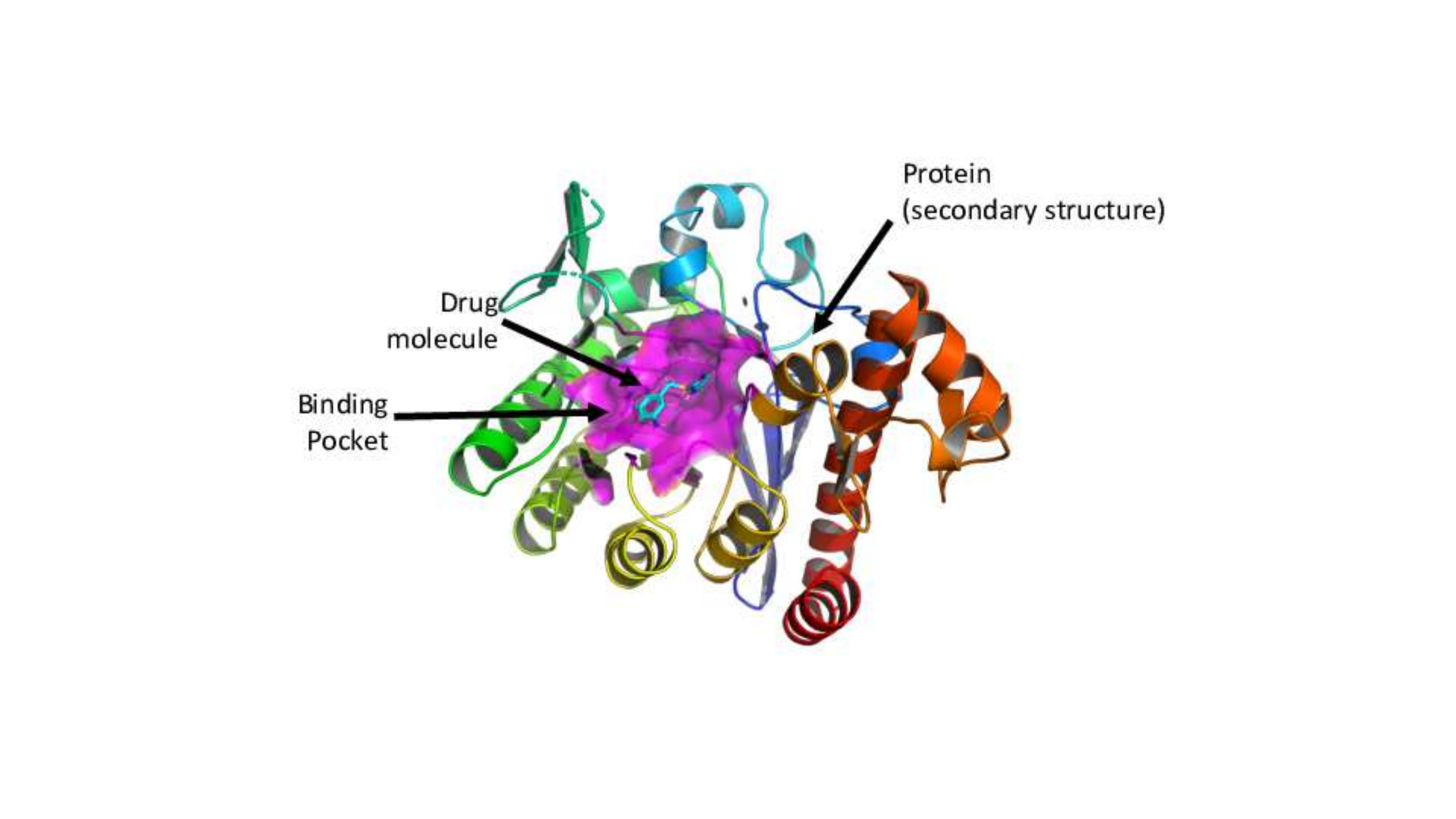}

  \caption{An example of a ligand-receptor complex with PDB code 1q63 from the PBDBbind database}
  \label{fig:binding_complex}
\end{figure*}

\section{Materials and Methods}
\subsection{Data}

The PDBBind database, a curated subset of the Protein Data Bank (PDB)~(\citealt{burley_protein_2019}) and initially developed for use in molecular dynamics pipelines, is a popular choice for the development of machine learning based scoring functions~(\citealt{feinberg_potentialnet_2018,jimenez_kdeep:_2018}) as it represents the largest publicly available source of experimentally determined molecular binding structures (e.g. protein-protein, protein-ligand, etc.). We consider only protein-ligand binding structures in this study. The database is essentially organized into two subsets (\textit{general} and \textit{refined}) based upon criteria that consider the nature of the complex (e.g. complexes that have ligands with molecular weight above 1000\si{\dalton}~are excluded from refined), the quality of the binding data (e.g. complexes with $\textit{IC}_{50}$ but no $k_{i}$ or ${k_d}$ measurement are not included in refined) and the quality of the complex structure (e.g. resolution of crystal structure must be better than 2.5\si{\angstrom}). From the refined set, a \textit{core} set is compiled to provide a representative set for validation using a clustering protocol. 

The 2016 edition of PDBBind used in this study, consists of 13,308 protein-ligand binding complexes in general, 4,057 complexes in refined, and 290 complexes in core.  An example input is shown in Figure \ref{fig:binding_complex}.

All docking complex data is generated using the in-house developed ConveyorLC toolchain~(\citealt{zhang_message_2013,zhang_toward_2014}), where docking poses are generated using the Vina scoring function~(\citealt{trott_autodock_2010}) and the top 10 poses are re-scored with the molecular mechanics/generalized Born surface area (MM/GBSA) method. A relative higher solute (or interior) dielectric constant (i.e. $\epsilon = 4$) is used in the MM/GBSA rescoring since previous studies demonstrate remarkable improvement in the pose ranking~(\citealt{sun_assessing_2014}). MM/GBSA re-scoring is about an order of magnitude more computationally costly than docking functions but have shown the potential to improve the accuracy of docking results while still being several orders of magnitude faster than more costly molecular dynamics simulations~(\citealt{zhang_toward_2014}).

\subsubsection{Pre-processing}
Before extracting the respective three-dimensional representations for each deep learning model, a common pre-processing protocol was applied to the binding complex structures provided in Protein Data Bank (.pdb) format by the PDBBind database. The process closely mirrors~\citealt{stepniewska-dziubinska_development_2018} to support a reproducible and comparable pipeline. 

All protein-ligand binding complexes were protonated and charges are solved using UCSF Chimera~(\citealt{pettersen_ucsf_2004}) with AMBER ff14SB~(\citealt{maier_ff14sb_2015}) for standard residues and AM1-BCC~(\citealt{jakalian_fast_2000}) for non-standard residues, the default settings for the program. No additional steps were taken for crystal structure data. For docking structures, water molecules were removed from the hold out PDBBind 2016 \textit{core} set to simulate realistic conditions of evaluating new docking poses. The results of this protocol produces a Tripos Mol2 (.mol2) file for each protein pocket.

\subsubsection{Feature Extraction}\label{subsec:feature}
A common atomic representation based on that of~\citealt{stepniewska-dziubinska_development_2018} was used for input to the structure based deep learning models. We consider only the heavy atoms from each biological structure and heteroatoms (i.e. Oxygens from crystallized water molecules).
\begin{itemize}
    \item Element type: one-hot encoding of  \textit{B},  \textit{C},  \textit{N},  \textit{O},  \textit{P},  \textit{S}, \textit{Se}, \textit{halogen} or \textit{metal} 
    \item Atom hybridization (1, 2, or 3) 
    \item Number of heavy atom bonds (i.e. heavy valence) 
    \item Number of bonds with other heteroatoms 
    \item Structural properties: bit vector (1 where present) encoding of \textit{hydrophobic}, \textit{aromatic}, \textit{acceptor}, \textit{donor}, \textit{ring} 
    \item Partial charge 
    \item Molecule type to indicate protein atom versus ligand atom (-1 for protein, 1 for ligand) 
    \item Van der Waals radius 
\end{itemize}

The OpenBabel cheminformatics tool (version 2.4.1)~(\citealt{oboyle_open_2011}) was used to extract the features for all binding complexes. Finally, all atomic coordinates are centered by each ligand to produce the spatial representation of the binding complex. 

\subsubsection{Training, Validation \& Testing Partitioning}
 The refined set was subtracted from the general set, and the core set was subtracted from the refined set such that there are no overlaps between the three subsets. We hold out the core set to be used as testing data, keeping the remaining general and refined complexes as training data. Due to the relatively small size of our datasets as compared to those in other domains such as computer vision~(\citealt{deng_imagenet_2009}), we account for potential bias in validation splits. For the general \& refined sets, we compute binding affinity quintiles of each set independently, then sample 10\% of the data from each quintile to form each validation set for general and refined. 

\subsection{3D-CNN}
\begin{figure*}
    \centering
    \includegraphics[width=0.78\linewidth]{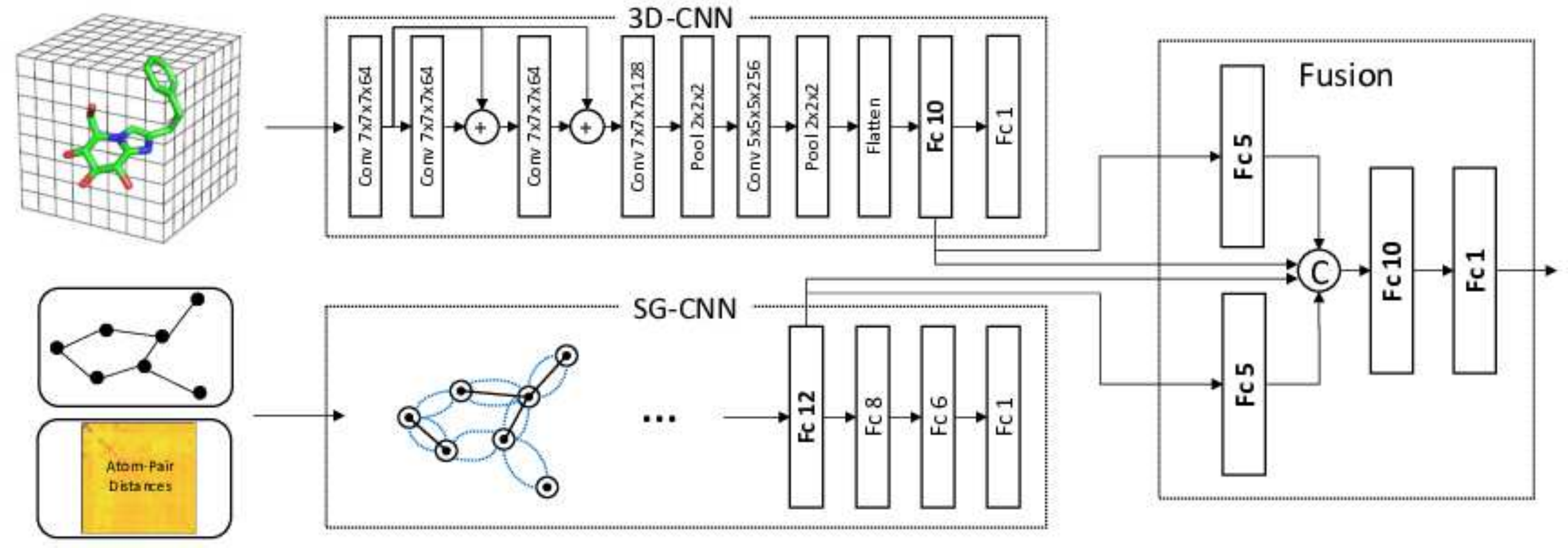}
    \caption{The proposed mid-level fusion model together with 3D-CNN and SG-CNN.}
    \label{fig:fusion}
\end{figure*}

3D-CNNs have been widely used in a variety of computer vision applications such as 3D image segmentation for medical diagnosis and video gesture/activity recognition. Recently, several 3D-CNN methods have been proposed for binding affinity prediction and other protein-ligand interactions in the drug discovery domain~(\citealt{wallach_atomnet:_2015,ragoza_proteinligand_2017, jimenez_deepsite:_2017, jimenez_kdeep:_2018, kuzminykh_3d_2018}). These methods are designed to capture 3D atomic features with their implicit interaction using 3D volume representations where atoms and their features are voxelized into the 3D voxel grid.

The 3D atomic representation used in this paper is described as follows. The input volume dimension is $ N \times N \times N \times C $ where $N$ is the voxel grid size in each axis (48 in our experiment) and $C$ is the number of atomic features described in Subsection \ref{subsec:feature} (19 in our experiment). The volume size in each dimension is approximately $48\si{\angstrom}$ where each voxel size is $1\si{\angstrom}$, which is sufficient to cover the entire pocket region while minimizing the collisions between atoms. Each atom is assigned to at least one voxel or more, depending on its Van der Waals radius or the user-defined size. In the case of the collisions between atoms, we apply element-wise addition to the atom features. Once all atoms are voxelized, Gaussian blur with $\sigma=1$ is applied in order to populate the atom features into neighboring voxels, similar to \citealt{kuzminykh_3d_2018}. 

The 3D-CNN used in this paper consists of 5 convolutional layers with two residual blocks, as illustrated in Figure~\ref{fig:fusion} (\textit{Top}). The residual short connection proposed in ResNet \citealt{he_deep_2016} allows the network to pass gradients to the next layers without non-linear activation. While it was originally designed for extremely deep layers, we observed that two residual blocks in the proposed 3D-CNN improve the prediction performance, compared to the one without a residual block and the previous models such as AtomNet~(\citealt{wallach_atomnet:_2015}). In addition, batch normalization is applied to individual convolutional layers to normalize each feature output across a mini-batch. We use the ReLU activation for nonlinearity.

\subsection{SG-CNN}
Deep learning approaches for modeling chemical graphs have demonstrated viability for learning continuous vector representations of molecular data as well as for property prediction tasks~(\citealt{duvenaud_convolutional_2015, gilmer_neural_2017, kearnes_molecular_2016}). The molecular graph considers atoms as nodes and bonds as edges between the nodes. The aforementioned 2D representation may be suitable for the modeling of chemical graphs however it fails to capture non-covalent interactions (e.g. atomic interactions between the ligand and the protein binding pocket) that are necessary to model more complex biological structures. Atomic Convolutional Neural Networks (ACNNs) were introduced in the work by~\citealt{gomes_atomic_2017} and allow for the specification of a ``local'' neighborhood for each atom that is based on euclidean distance, effectively relaxing the covalency requirement to form edges in the graph representation of a protein-ligand binding complex. The PotentialNet architecture presented by~\citealt{feinberg_potentialnet_2018} refines this idea by allowing for an arbitrary number of edge types and applies specialized update rules that are a generalization of Gated Graph Sequence Neural Networks~(\citealt{li_gated_2017}).

We adopt a spatial graph representation~(\citealt{feinberg_potentialnet_2018}) for the molecular complexes where the atoms in a given complex are considered as the \textit{nodes} within this graph representation. Both covalent and non-covalent bonds are represented through the use of a square $N\times N$ adjacency matrix $A$ and an $N\times M$ node feature matrix where $A \in \mathbb{R}^{N \times N}$ and $A_{ij}$ is equal to the euclidean distance (in angstroms~\si{\angstrom}) of atom $i$ and atom $j$. To further expand this representation as a 3D tensor, we define two thresholds for covalent and non-covalent ``neighborness'', $\alpha_{c}$ and $\alpha_{nc}$ respectively s.t. $A_{ij,c} = 0$ if $A_{ij} \geq \alpha_c$ and $A_{ij,nc} = 0$ if $A_{ij} \geq \alpha_{nc}$. In the context of this paper the thresholds used were $\alpha_{c}=1.5$\si{\angstrom} and $\alpha_{nc}=4.5$\si{\angstrom}. We found these settings lead to the best performance on the validation set. We implement the SG-CNN using the PyTorch Geometric (PyG) python library~(\citealt{fey_fast_2019}).

\subsection{Fusion}
Fusion models to combine multiple input sources or different feature representations have been applied to a number of computer vision applications, especially in the presence of multi-modal images or different image sensors. These fusion models benefit from multiple feature representations which are considered complementary to each other. In addition, fusion-based approaches increase robustness by reducing uncertainty of each feature representation or modality. Inspired by that, we propose to use a separate fusion neural network to combine feature representations from two independently trained models (3D-CNN and SG-CNN), each of which has its own strength and weakness. Such heterogeneous feature representations that the two models capture can enrich the proposed fusion model's features, which has strong potential to improve the performance of binding affinity prediction.

Among several ways to fuse models addressed in~\citealt{roitberg_analysis_2019}, we adopt mid-level to late fusion approaches. To the best of our knowledge, this fusion model is the first attempt to combine multiple model representations for the task of protein-ligand binding affinity prediction. One can add more features from different machine learning models into the fusion model for better prediction. 

Figure~\ref{fig:fusion} illustrates the proposed mid-level fusion model that combines two features from 3D-CNN and SG-CNN. For the input features in the fusion model, we use the second- and fourth-last layer's output activations from 3D-CNN and SG-CNN, respectively. Then each input feature is passed by a fully connected layer, and the outputs are concatenated with the original input features for cross-layer connections, before the next fully connected layer. We observed that this concatenation outperformed addition and the one without any connection. Similar to 3D-CNN, we apply batch normalization and ReLU-based non-linearity in each fully connected layer. Jointly training a fusion model including both CNN models is practically infeasible since training the entire model requires a significant amount of GPU memory and time. It is often required to use extremely small mini-batch sizes or to utilize a distributed model using multiple GPUs. Independently trained CNN models offer flexibility to use different fusion strategies without re-training the individual CNN models. Therefore, given separately trained CNN models, we performed forward propagation (prediction) to obtain the layer  output activations using all training, validation and test datasets. The output activations from the CNN network models become the inputs of the fusion model. Then we perform the fusion model training and validation, followed by testing.  It is important to note that training of the fusion model is done exclusively on the training and validation sets used to train the individual models and fusion training is never conducted on any of the held out test sets.
In addition to the mid-level fusion model above, we employ the late fusion approach. This fusion is carried out by averaging the final predictions of the CNN models. This approach is simple, but effective to combine multiple model predictions. We report the results using this method as well as those using the mid-level fusion approach. 

\subsection{Structure based clustering}
Protein-ligand complex binding pockets from the PDBBind database were compared to identify local regions surrounding ligands and perform structure-based clustering for evaluation of machine learning model performance. Clustering of the structures was performed using LGA~(\citealt{zemla_lga_2003}) on a whole-protein level as well as specific local substructures selected to represent ligand binding site regions.
The implemented approach can be briefly described as follows: for each protein-ligand complex, the binding site local environment was delineated using an initial 12.0\si{\angstrom} radius sphere centered at ligand atoms. The sphere size of 12.0\si{\angstrom} radius was selected in order to capture as much conformation information around the local environment as possible to allow detection of similarities between pockets even with different sizes of observed ligands. Previous research indicated that distances of 7.5\si{\angstrom} are an upper limit in capturing informative functional properties for clustering purposes~(\citealt{yoon_clustering_2007}), so in our approach to enhance detection of functional residues, we additionally collected information on protein-ligand interface residues identified within the distance of 4.0\si{\angstrom}. Defined this way ligand-pocket templates were used to search for structure similarities with any local region from all PDBBind complexes (all against all search). Structure similarity between superimposed aligned regions was measured using the GDC metric~(\citealt{keedy_other_2009}) which evaluates similarity based on conformation of all atoms from compared substructures, not only Calpha positions. For all of 4,464 protein-ligand complexes for which binding affinity scores were provided in the PDBBind (release 2018) refined set an “all against all” matrix of similarity scores was created, and 830 clusters were formed hierarchically using Euclidean distance.  For evaluation, only in release 2016 were included.

\section{Results}
The PDBBind 2016 core dataset was used to evaluate the following hypotheses: 

\begin{itemize}
    \item The two CNN models provide complimentary information.
    \item The fusion model learns to integrate the two CNN models and improve prediction over the individual ones.
    \item The machine learning models retain prediction accuracy when presented with docked poses rather than crystal structures.
    \item The machine learning models are as accurate as the more computationally costly MM/GBSA re-scoring function.
\end{itemize}

\begin{table}
\centering
\begin{threeparttable}
\caption{Performance of Binding affinity prediction on the crystal structure of PDBBind 2016 core set. \textit{Top}: comparison of the proposed fusion approaches with individual and existing models. \textit{R}: refined set, \textit{G}: general set.}

  \begin{tabular}{|l|l|l|l|l|l|l|}
  \hline
    Model & $\textit{r}^{2}$ & $\textit{Pearson r}$ & $\textit{Spearman r}$ & $\textit{MAE}$ & $\textit{RMSE}$ \\ \hline\hline
    SG-CNN (R) &.424 & .666 & .647 & 1.321 & 1.650 \\ \hline
    SG-CNN (G) &.519 & .747 & .746 & 1.194 & 1.508 \\ \hline
    SG-CNN (R + G) & .600 & .782 & .766 & 1.084 & 1.375 \\ \hline
    3D-CNN (R) & .523 & .723 & .716 & 1.164 & 1.501 \\ \hline
    3D-CNN (G) & .420 & .649 & .658 & 1.294 & 1.655 \\ \hline
    3D-CNN (R + G) & .397 & .677 & .657 & 1.334 & 1.688 \\ \hline
    Late Fusion & .628 & .808 & .803 & 1.044 & 1.326 \\  \hline
    Mid-level Fusion & \textbf{.638} & \textbf{.810} & \textbf{.807} & \textbf{1.019} & \textbf{1.308} \\ \hline
    $\text{Pafnucy}^{a}$ & - & 0.78 & - & 1.13 & 1.42 \\ \hline
  \end{tabular}

\begin{tablenotes}\footnotesize
    \item[a] \citealt{stepniewska-dziubinska_development_2018}
\end{tablenotes}
\end{threeparttable}
\end{table}

\begin{table}
\centering
\begin{threeparttable}
\caption{Comparison of the proposed mid-level fusion model with physics-based scoring functions on the crystal structures of PDBBind 2016 core set. We give the results for the 243 complexes for which it was possible to compute a score across all methods. The correlation coefficients of Vina and MM-GBSA scoring functions are given as absolute values.}
  \begin{tabular}{|l|l|l|l|l|l|l|l|}
  \hline
    Method & $\textit{Pearson r}$ & $\textit{Spearman r}$ & $\textit{MAE}$ & $\textit{RMSE}$ \\ \hline\hline
    Vina & .599 & .605 & - & - \\ \hline
    MM/GBSA & .647 & .649 & - & - \\ \hline
    Mid-level Fusion & \textbf{.803} & \textbf{.797} & \textbf{1.035} & \textbf{1.327} \\ \hline
  \end{tabular}
  \label{tab:results_crystal}
 \end{threeparttable}
\end{table}

\subsection*{Prediction Performance on PDBBind-2016 Crystal Structure}

Table~\ref{tab:results_crystal} summarizes the model performance on the crystal structure of PDBBind 2016 \textit{core} set. Training on both PDBBind's general and refined data were considered. While training on the larger \textit{general} dataset could improve performance, it has the drawback of noisier binding affinity measurements and lower resolution 3D structures (typically larger than 2.5\si{\angstrom})~(\citealt{su_comparative_2019}).  Table ~\ref{tab:results_crystal} shows that the SG-CNN model trained on both the general and refined sets (SG-CNN (general + refined)) outperformed the other SG-CNNs. For the 3D-CNN, however, the 3D-CNN model trained on the refined set only (3D-CNN (refined)) achieved the highest validation and test accuracy. The performance difference between 3D-CNN and SG-CNN offers several technical insights of the model behaviors. First, SG-CNN models benefit from more training samples while they are less sensitive to the structural resolution. On the other hand, the lower structural resolution of the complexes in the general set significantly degrades the performance of the 3D-CNN models because incorrect atom positions larger than 2.5\si{\angstrom} cause more than 2 voxel deviation in the current 3D representation. The fusion model used the SG-CNN (general + refined) and 3D-CNN (refined) as the input features.  Among all results, the proposed mid-level fusion approach outperformed the individual models. The late fusion (averaging) also achieved a higher accuracy than the individual models while the performance difference between the mid-level and late fusion models is marginal.  In addition, the results show that the fusion model has substantially higher Pearson correlation coefficient than MM-GBSA scoring (0.803 versus 0.647).

\subsection*{Prediction Performance on PDBBind-2016 Docking Poses}

\begin{table}
\centering
\begin{threeparttable}
\caption{Performance on PDBBind 2016 Core Set - Docking Poses.}
  \begin{tabular}{|l|l|l|l|l|l|l|}
  \hline
    Model & $\textit{Pearson r}$ & $\textit{Spearman r}$ & $\textit{MAE}$ & $\textit{RMSE}$ \\ \hline\hline
    SG-CNN & .656 & .625 & 1.343 & \textbf{1.649} \\ \hline
    3D-CNN & .523 & .503 & 1.843 & 2.358 \\ \hline
    Vina  & .616 & .618 & - & - \\ \hline
    MM-GBSA & .629 & .641 & -  & - \\ \hline
    Late Fusion & \textbf{.685} & \textbf{.668} & \textbf{1.34}  & 1.701 \\ \hline
    Mid-level Fusion & .677 & .647 & 1.351  & 1.715 \\ \hline
  \end{tabular}
  \label{tab:results_docking}
 \end{threeparttable}
\end{table}

Scoring the binding affinity of a crystal complex is useful for separating the scoring task from the ligand pose selection problem. In practice, however, the correct ligand pose will not be known and the scoring function will evaluate noisier and error prone docking poses.  To address this problem, the machine learning models scored the top 10 Vina poses and report the highest binding affinity for each complex for 257 test complexes where MM/GBSA re-scoring calculations completed.  Waters captured in the crystal structure were removed since this information can artificially constrain the docking poses and inflate performance. A modest increase in performance was observed when water is retained (results not shown). Prediction performance on docking poses is summarized in Table \ref{tab:results_docking} and compared with the original Vina docking score and the more expensive MM/GBSA re-scoring function. As expected, overall performance decreases relative to scoring crystal structures, which can in part be explained by fewer correct poses to re-score.  Using a maximum RMSD of 2\si{\angstrom} threshold between a docked pose and the crystal pose, a correct pose is found among the top 10 Vina poses in only 77\% of the cases when evaluating the refined dataset. Nonetheless, the fusion model's Pearson correlation coefficient, remains higher than the computationally costly MM/GBSA scores and Vina scores (0.685 versus 0.629 and 0.616), motivating use of the Fusion model over the scoring functions.

Classification performance was evaluated for predicting non-binders (threshold set to pKi/pKd $<$ 5) and binders (threshold set to pKi/pKd $>$ 8).  Detecting promiscuous binders can be important for predicting toxicity while screening a large compound library against a single protein target.  Measuring both screening tasks provides information on how prediction scores at both extremes of the distribution inform bind/no-bind recognition.   The results are summarized in Table S1 and show ROC AUC values of 0.82 and higher.

\section{Discussion}
\begin{figure}
  \centering
    \includegraphics[width=0.5\linewidth]{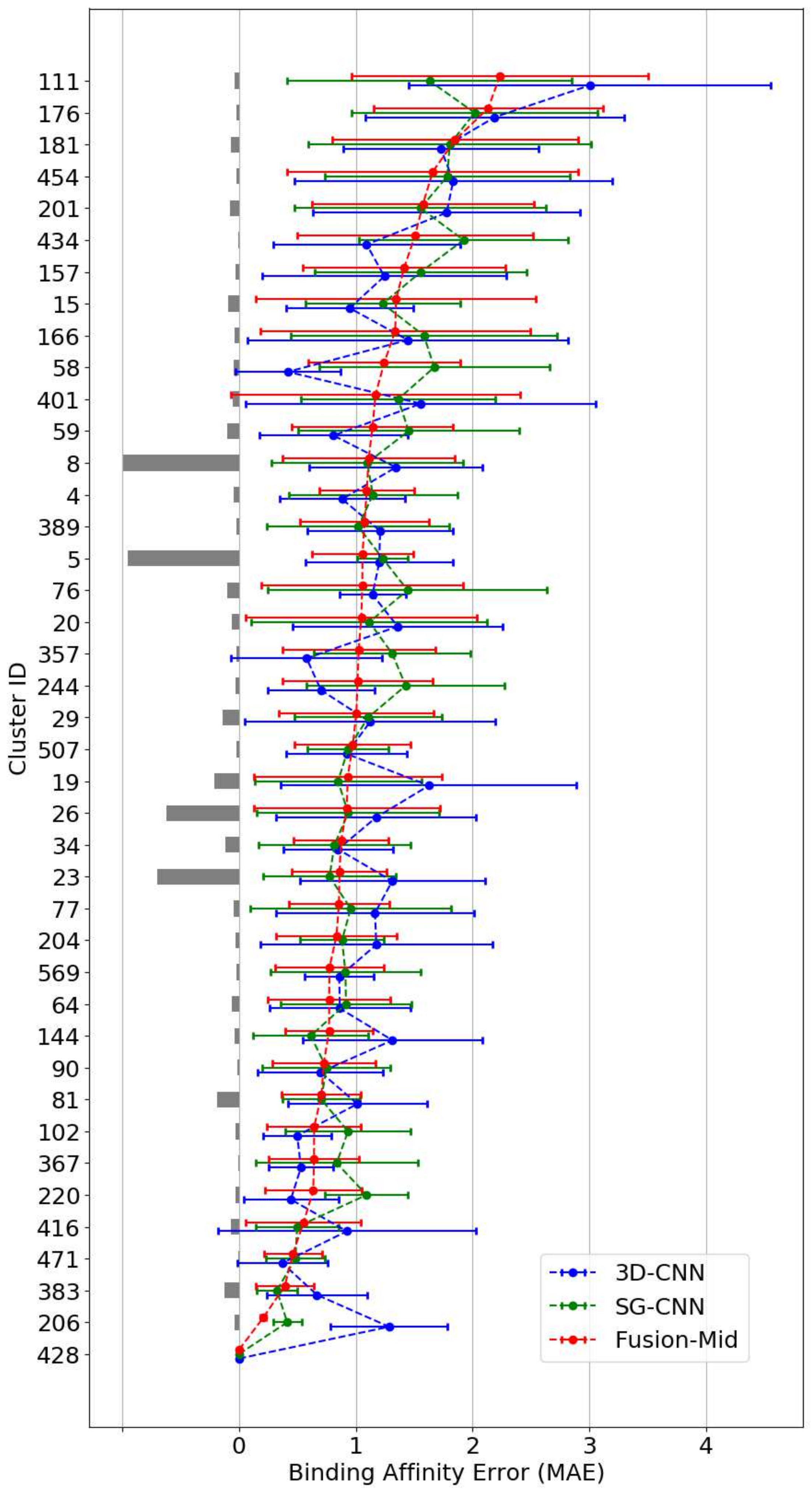}

  \caption{MAE (x-axis) with standard deviation for groups (y-axis) based on the pocket and the ligand positioning. MAE is shown for the machine learning models. The number of complexes in the \textit{refined} training set is shown for each cluster (\textit{gray} bars).}
  \label{fig:results_crystal_pocketcluster_mae}
\end{figure}

The complexes found in the evaluation set came from 41 of the 830 structure based clusters.  These clusters were used to assess prediction performance across the different clusters. (The complete listing of clusters is provided as a supplemental file.)  The MAE is shown in the Figure \ref{fig:results_crystal_pocketcluster_mae} and shows a trend of varying error exceeding 2 logs in some cases suggesting more accurate predictions for specific protein clusters.  

\begin{figure}
  \centering
    \includegraphics[width=0.5\linewidth]{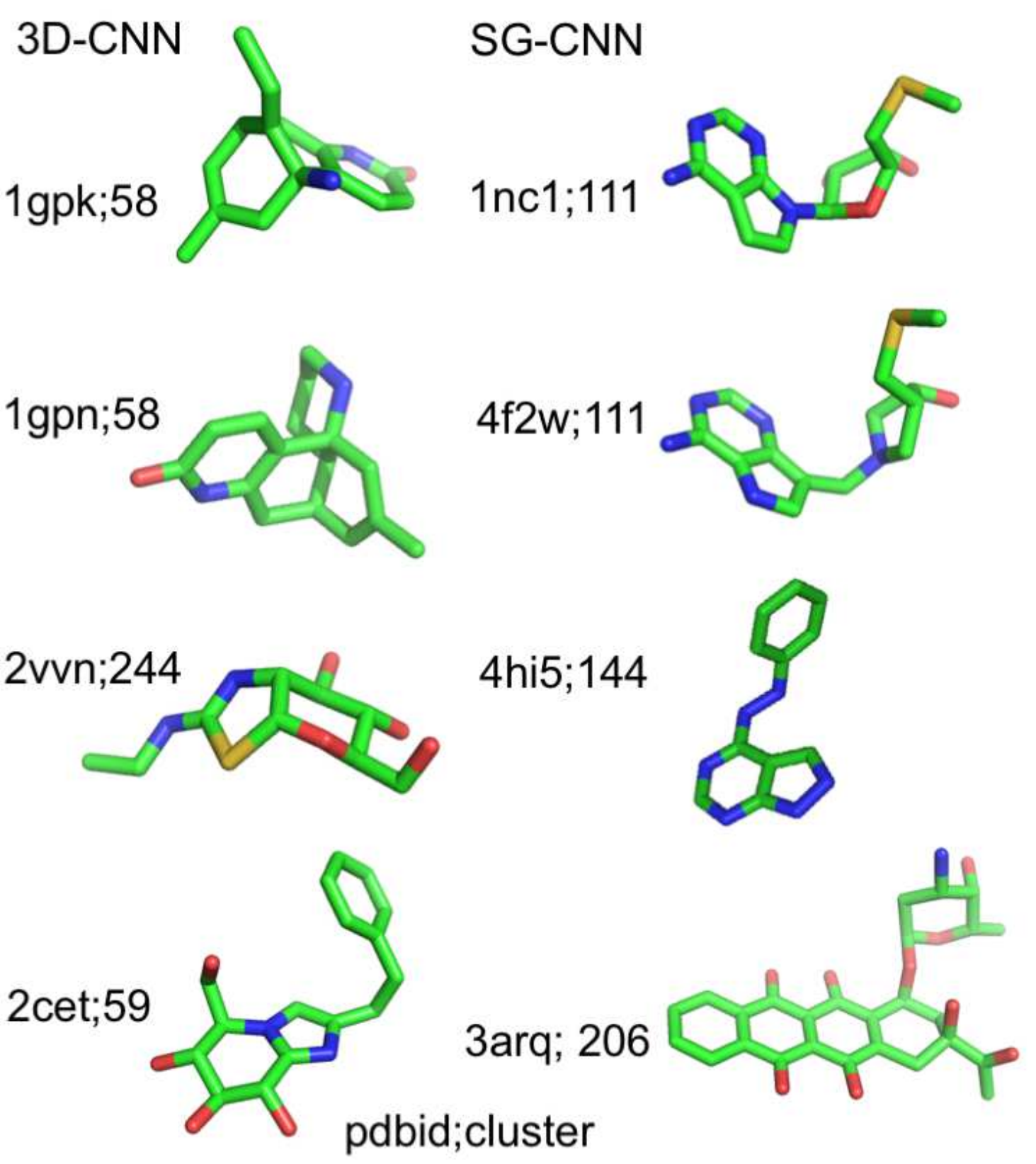}
    
  \caption{Structure for 8 compounds with maximum difference in prediction between two models.  Top 4 cases shown where error is lower for 3D-CNN or SG-CNN.  Hydrogens are not shown, images are generated with Pymol~(\citealt{schrodinger_llc_pymol_2015})}
  \label{fig:example_compounds}
\end{figure}

SMILES strings were constructed for 269 of the 290 compounds referenced in the 2016 \textit{core} set.  51 compounds were found to occur in both the holdout set and the \textit{refined} training set.  Tanimoto distance  between each test ligand and its most similar ligand in the training set from the same cluster was compared with MAE but no correlation was found.  These results suggest that models are learning the structurally important features, but other chemical and physical information may be needed.   There are six clusters (111,144,176, 20, 206 and 401) with at least two complexes within the respective cluster where a difference in error between the two models is at least 1 log and the SG-CNN does consistently better in each case.  Similarly, there are four clusters (244,58,59 and 64) where the reverse occurs and the 3D-CNN shows consistently lower error.  Figure \ref{fig:example_compounds} shows the top 8 compounds with maximum prediction discrepancy in the two models. It is still unclear whether there are important structural differences in these clusters that explain the advantage of one model over the other.   The first two examples in Figure \ref{fig:example_compounds} highlight compounds that interact with the same pocket type, 58 and 111 for the 3D-CNN and SG-CNN respectively.  There are many other clusters where neither  model has a clear advantage.  Nevertheless, the models clearly exhibit distinct performance profiles.  While the Fusion model exhibits better overall performance in more clusters than its constituent models, it is not able to give the lowest MAE in every cluster.

\section{Conclusion}
The results show that the two CNN models provide complimentary predictions for many test complexes.  The current SG-CNN implementation does not explicitly capture bond angles and we speculate that in some cases where the 3D shape of the molecule is important, the 3D-CNN may have an advantage.  On the other hand, the SG-CNN likely benefits from a more explicit representation of pairwise interactions, which leads to fewer parameters to learn.  The benefit of using both models is supported by the performance of the Fusion models, which yield improved overall performance compared to the individual models.  An area for future improvement could be in exploring activity maps such as those introduced in~\citealt{hochuli_visualizing_2018} to identify more explicitly, which aspects of the input features are most important in determining binding affinity.

Although prediction accuracy drops when using docked poses, the fusion models maintain improved accuracy over the docking and MM/GBSA scores.   The reduction in accuracy suggest that the machine learning models are more sensitive to mis-scoring incorrect poses.  An opportunity for improvement would be to add a machine learning task to classify correct and incorrect poses.  This would allow the model to learn the binding affinity value while directly training to differentiate correct and incorrect poses.  This could help filter out binding affinity scores that use poses predicted to be incorrect.  

The structure based clustering of pockets provides another way to  incorporate prior structural knowledge.  Prediction error differs for different cluster types, suggesting that some categories of binding interactions are harder to predict than others.  The clusters constructed for PDBBind can be used to classify new docked poses according to the existing clusters and recognize when new complexes fall outside of the current set of clusters.

The machine learning model prediction error appears to be surprisingly robust when predicting on new ligands in recognized pockets. Moreover,  accuracy should continue to improve as the amount of experimental data  grows.   We conclude that the Fusion model will become a more computationally efficient alternative to the MM/GBSA re-scoring function.

\section*{Acknowledgements}

None.

\section*{Funding}
This work was supported by American Heart Association Cooperative Research and Development Agreement TC02274. This work was performed under the auspices of the U.S. Department of Energy by Lawrence Livermore National Laboratory under Contract DE-AC52-07NA27344. LLNL-JRNL-804162.

\vspace*{-12pt}

\bibliographystyle{natbib}

\bibliography{atom_computing}

\end{document}


\maketitle

\subsection*{Spatial Graph Convolutional Network Architecture}

The Spatial Graph Convolutional Neural Network (SG-CNN) presented here is composed of a number of smaller neural networks and can be described in terms of a \textit{propagation} layer, a \textit{gather} or aggregation step, and finally, an output fully-connected network. The \textit{propagation} layer can be understood as the portion of the network where messages are passed between the atoms and used to compute new feature values for each node, with a total of $k_{i}$ rounds of message passing, where $i$ (wlog) corresponds to the ${i^{th}}$ propagation layer.


Using this architecture, we define two distinct propagation layers for covalent and non-covalent interactions. In both cases, we use a single scalar value for the edge feature, the euclidean distance (measured in Angstroms, \si{\angstrom}) between nodes $v_i$ and $v_j$. Propagation is performed for $k=2$ rounds for both cases, then the attention operation is performed on the final propagation output. For covalent propagation, the input node feature size is 20 and the output size is 16 after the attention operation is applied. For non-covalent propagation, the node features that are the result of the covalent attention operation are used as the initial feature set, the output node feature size of this layer is 12. The resulting features of the non-covalent propagation are then ``gathered'' across the ligand nodes in the graph to produce a ``flattened'' vector representation by taking a node-wise summation of the features. This graph summary feature is then fed through the output fully-connected network to produce the binding affinity prediction $\hat{y}$.

\begin{gather*}
    \forall i~h^{k=0}_{i} = x_i\qquad \text{for $k=0$} \tag{intialization of node features}
\end{gather*}

\begin{gather*}
    h_{i}^{k} = GRU(h_{i}^{k-1}, \sum_{j \in \mathcal{N}^{(e)}(v_{i})}f_{e}(h_j))\qquad\text{for $k \in \{1...K\}$} \tag{message passing}
\end{gather*} 

where $x_{i}$ is the feature vector of node $v_{i}$.

\begin{gather*}
    f_{e}(h_i) = \Theta h_{i}+\sum_{j \in \mathcal{N}(i)}h_{j}\cdot f_\theta(e_{i,j}) \tag{outer edge network}
\end{gather*}

where $\Theta$ is a learned set of parameters, $x_i$ is the set of features for $n_i$, and $f_\theta$ is a neural network that computes a new edge feature for the edge $e_{i,j}$ between $v_i$ and $v_j$.

\begin{gather*}
    f_{\theta}(e_{i,j}) = \phi(f_{\theta_1}(\phi(f_{\theta_2}(e_{i,j})))) \tag{inner edge network}
\end{gather*}

where $f_{\theta_0}$ and $f_{\theta_1}$ are neural networks and $\phi$ is the Softsign~(\citealt{turian_quadratic_2009}) activation (defined as $1/1 + \vert x \vert$).

\begin{gather*}
    h^{K^{\prime}} = \sigma(p(h^{K}, h^{k=0})) \odot \big(q(b^{K})) \tag{attention operation}
\end{gather*}

where $\sigma$ is the softmax activation function, $q$ and $p$ are neural networks.

\begin{gather*}
    h_{gather} = \sum_{v \in G_{ligand}} h_v \tag{gather step}
\end{gather*}

where $G_{ligand}$ is the ligand subgraph of the binding complex graph.

\begin{gather*}
    \hat{y} = g(h_{gather}) = g_2(ReLU(g_1(ReLU(g_0(h_{gather}))))) \tag{output network}
\end{gather*}

\newpage 
\subsection*{Experiment Setup}
The 3D-CNN, SG-CNN, and the fusion network models, used an Adam optimizer with learning rate of $1\times10^{-3}$, $1\times10^{-3}$, and $1\times10^{-3}$, respectively. The mini-batch sizes are 50, 8, 100, and the approximate number of epochs are 200, 200, 1000, respectively. We observed that smaller mini-batch sizes drastically improved the model performance in the training of the SG-CNN model while the effect of different mini-batch sizes was negligible in 3D-CNN and the fusion model.

The 3D-CNN and fusion models were developed using the Tensorflow python library~(\citealt{abadi_tensorflow_nodate}). The SG-CNN was developed using the PyTorch and PyTorch Geometric (PyG)python libraries~(\citealt{paszke_pytorch_2019, fey_fast_2019}) .

\newpage

\begin{figure*}[ht!]
  \centering
    \includegraphics[width=0.8\linewidth]{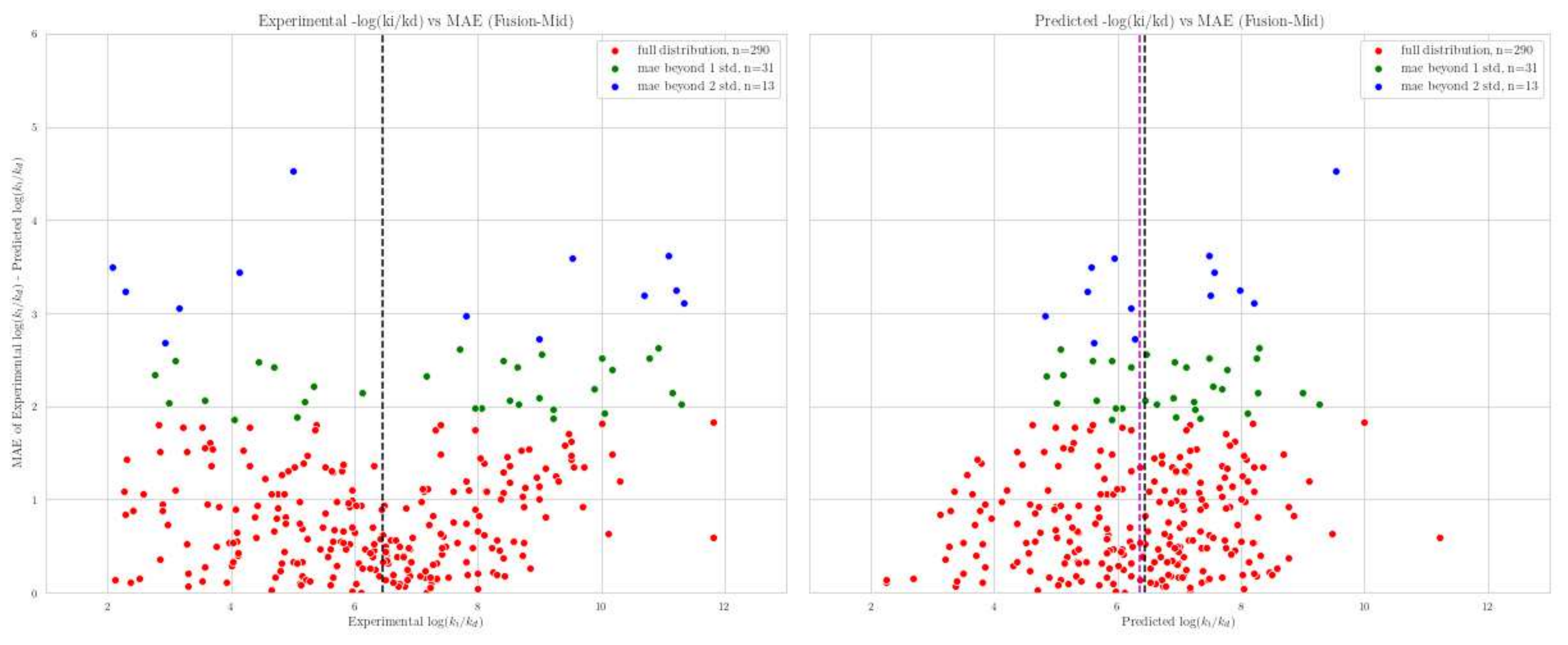}
  \caption{}
  \label{fig:results_crystal_casf2016_mae2}
\end{figure*}

\paragraph{Visualizing Model Bias and MAE distribution}

\begin{figure}[ht!]
    \centering
    \includegraphics[width=0.95\linewidth]{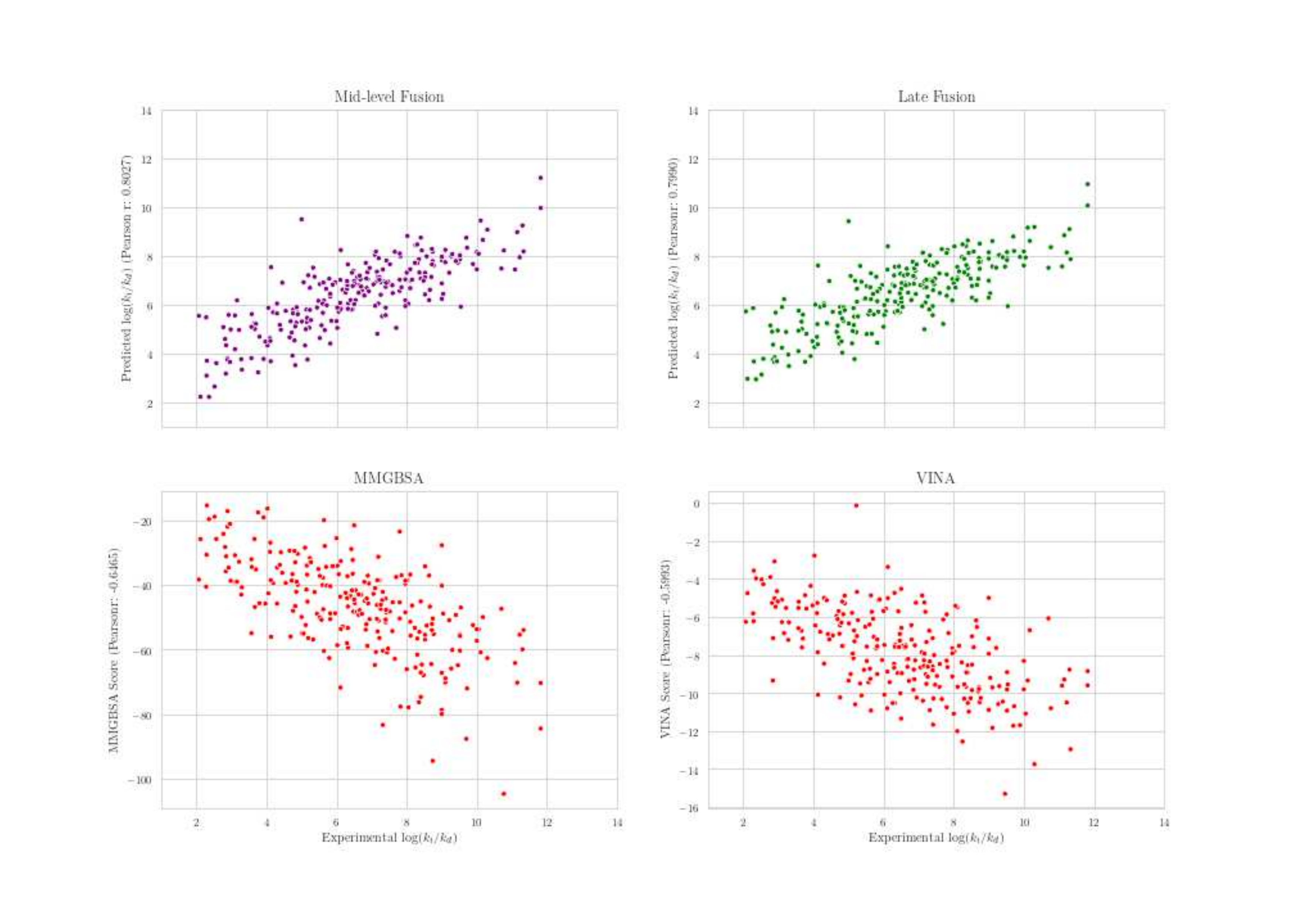}
    \caption{Scatter plots for each scoring method and the experiment binding affinity measurement.}
    \label{fig:results_crystal_scatter}
    
\end{figure}

In figure~\ref{fig:results_crystal_scatter} the PDBBind 2016 Core set is color coded into 3 groups according to the MAE of the experimental $log(k_i/k_d)$ (left panel) and the predicted $log(k_i/k_d)$ (right panel).  Predictions that fall below one standard deviation of the mean MAE value for a given model (red), between 1-2 standard deviations (green), and exceed 2 standard deviations (blue). Figure \ref{fig:results_crystal_scatter} shows that models predict over the full range of binding affinity values, however, predictions are biased toward the mean. As the error increases (higher values on the y-axis), the left panel plot shows how high error predictions have trouble predicting the tails of the distribution. The right panel plot shows that the models predict closer to the mean in cases of high error. The black vertical line gives the location of the ground truth mean and the purple vertical line gives the predicted mean.

\newpage

\paragraph{Relationship between scoring function output with experimental measurement.} Figure~\ref{fig:results_crystal_scatter} gives scatter plots of the scores from Mid-level Fusion, Late Fusion, MM/GBSA, and Vina scoring methods versus the experimental binding affinity for the 242 complexes of the 2016 core set for which a score across all methods was possible to obtain. Both Fusion methods show a significant improvement over physics-based scoring methods (in terms of Pearson correlation coefficient) with the experimental $log(k_i/k_d)$.

\newpage

\begin{figure*}[ht!]
  \centering
    \includegraphics[width=0.95\linewidth]{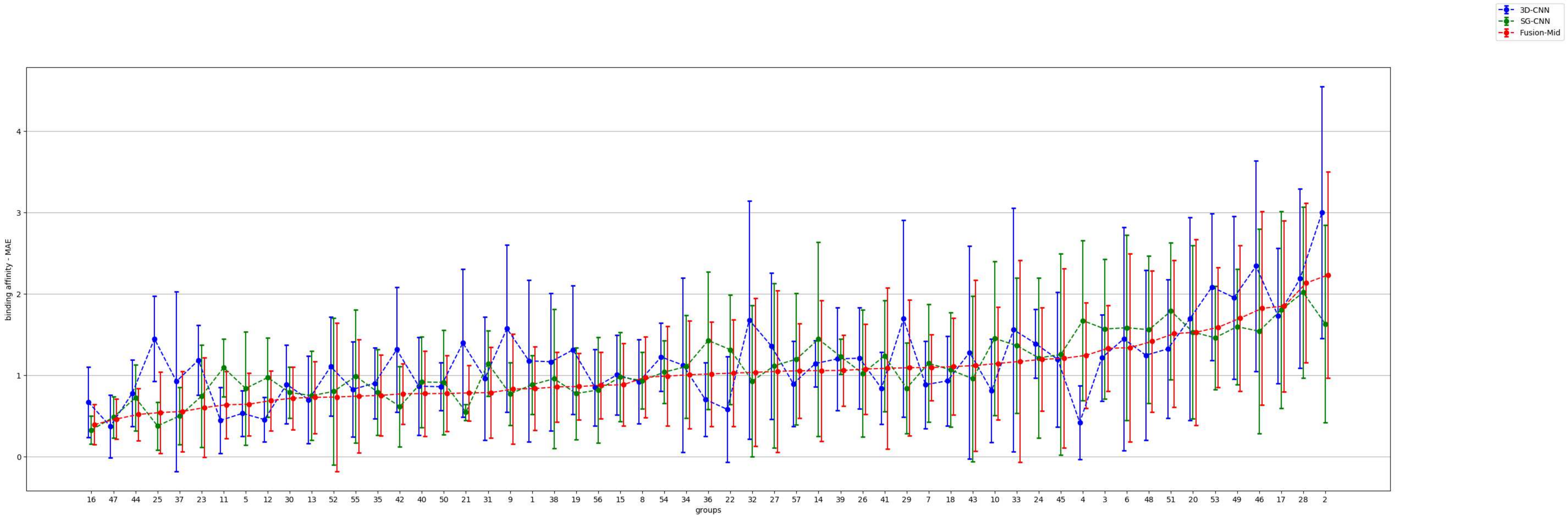}
  \caption{MAE (y-axis) with standard deviation for 57 functional groups (x-axis) defined by CASF-2016. MAE is shown for the machine learning models.}
  \label{fig:results_crystal_casf2016_mae}
\end{figure*}
\paragraph{Performance on protein targets (CASF-2016)} To consider prediction performance of the different models, mean absolute error (MAE) is shown for the 57 functional categories defined by the Comparative Assessment of Scoring Functions (CASF-2016) complexes, which consists of 285 PDBBind \textit{core} complexes with 5 complexes per category.  The results are shown in Figure \ref{fig:results_crystal_casf2016_mae} with the categories sorted on the x-axis by increasing Mid-fusion MAE. The figure shows significantly different Mid-fusion MAE between groups, ranging from approximately 0.5 to over 2 log units.  The protein categories are limited by manual curation and were not reported for the complete collection of PDBBind complexes, making it difficult to assess overlap between binding pockets found in the training data and the test set.  To address this, a binding pocket oriented structure based clustering scheme was applied to the complete collection of PDBBind complexes.

\newpage
'
\begin{table}[h!]
\centering
\caption{Comparison of classifier performance on PDBBind 2016 Core Set - Docking Poses. SD=standard deviation.}
  \begin{tabular}{|l|l|l|l|l|l|}
  \hline
    Model & $\textit{Bind ROC AUC}$ & $\textit{SD}$ & $\textit{No-bind ROC AUC}$ & $\textit{SD}$ \\ \hline\hline
    SG-CNN & .784 & .066 & .829 & .063 \\ \hline
    3D-CNN & .747& .081 & .774 & .063 \\ \hline
    Vina  & .788 & .071 & .848 & .052 \\ \hline
    MM/GBSA & \textbf{.828} & .064 & .833 & .057  \\ \hline
    Late Fusion & .82 & .065 & \textbf{.859} & .054 \\ \hline
    Mid-level Fusion & .806 & .07 & .853 & .055 \\ \hline
  \end{tabular}
  \label{tab:results_docking_classifier}
\end{table}

\paragraph{Classification of binders, performance comparison between fusion models and physics-based scoring}For the \textit{bind} detection screening task, the results are summarized as ROC AUC and reflect randomly partitioning the PDBBind 2016 \textit{core} set into 5 non-overlapping folds, computing the ROC AUC on each fold, repeating this procedure 100 times and taking the average. The datasets for both tasks maintain a similar class imbalance of 25\% positives and 75\% negatives.  Table \ref{tab:results_docking_classifier} show that MM/GBSA has slightly better performance than the Fusion model and both methods show a small improvement (0.04) over Vina.  For the \textit{no-bind} task, while the Fusion model had the highest ROC AUC the margin of difference was negligible compared to Vina (0.011). 

\newpage

\begin{figure}[ht!]
    \centering
    \includegraphics[width=0.95\linewidth]{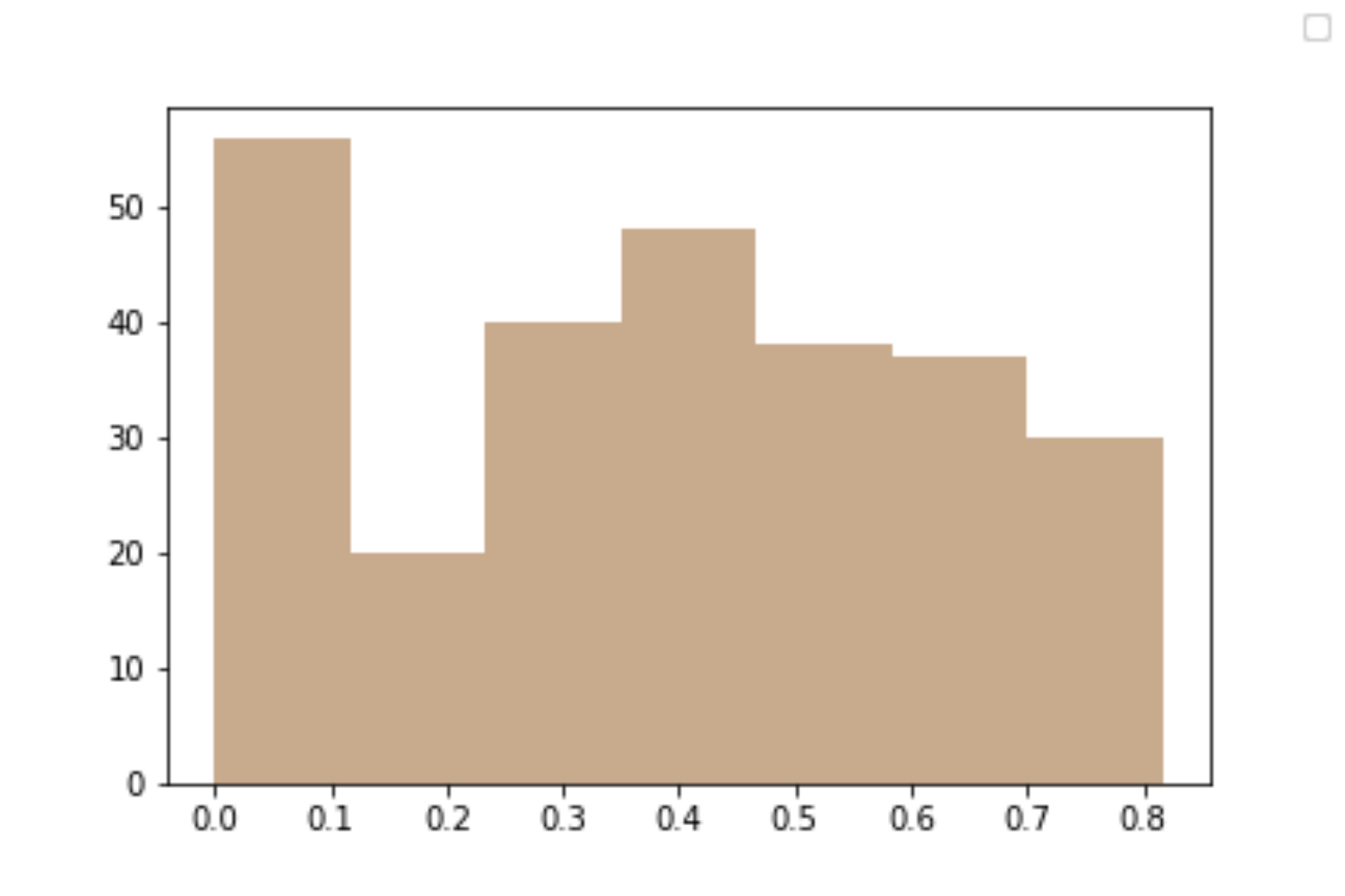}
    \caption{Histogram counting minimum Tanimoto distance between each compound in test and nearest match in PDBBind 20016 refined.}
    \label{fig:td_test_train_hist}
\end{figure}

\paragraph{Structure similarity between refined abd core sets of PDBBind 2016}In order to gain an understanding of how ``similar'' our training set (in terms of the refined set) was to our testing set (the core set), we consider structural similarity between the ligands in the refined and core sets as measure by the tanimoto distance metric. Figure~\ref{fig:td_test_train_hist} illustrates the distribution of the tanimoto distance between each ligand in the core set with its nearest neighbor in the refined set. 

\scriptsize{\subsubsection*{Disclaimer}
This document was prepared as an account of work sponsored by an agency of the United States government. Neither the United States government nor Lawrence Livermore National Security, LLC, nor any of their employees makes any warranty, expressed or implied, or assumes any legal liability or responsibility for the accuracy, completeness, or usefulness of any information, apparatus, product, or process disclosed, or represents that its use would not infringe privately owned rights. Reference herein to any specific commercial product, process, or service by trade name, trademark, manufacturer, or otherwise does not necessarily constitute or imply its endorsement, recommendation, or favoring by the United States government or Lawrence Livermore National Security, LLC. The views and opinions of authors expressed herein do not necessarily state or reflect those of the United States government or Lawrence Livermore National Security, LLC, and shall not be used for advertising or product endorsement purposes.
}

\newpage

\normalsize{\bibliographystyle{natbib}
\bibliography{atom_computing}}